\documentclass[conference]{IEEEtran}
\pdfoutput=1
\usepackage{etex}

\usepackage[utf8]{inputenc}

\usepackage{float}
\usepackage{varioref}

\usepackage[utf8]{inputenc}

\usepackage{alltt}
\usepackage{palatino}

\usepackage[english]{babel}
\usepackage{verbatim} 
\usepackage{tabularx} 
\usepackage{listings} 
\usepackage{graphicx}

\usepackage{rotating}
\usepackage{balance}
\usepackage{numprint}
\npthousandsep{,}
\usepackage{multirow}
\usepackage{framed}
\usepackage[usenames,dvipsnames]{xcolor}
\usepackage[]{mdframed}

\usepackage{algorithm}
\usepackage{algorithmic}

\usepackage{verbatim}
\usepackage[caption=false]{subfig}
\usepackage{xspace}
\usepackage{booktabs}
\usepackage{textcomp}
\usepackage{letltxmacro}
\usepackage[]{pdfcomment}

\newcommand*{\SavedLstInline}{}
\LetLtxMacro\SavedLstInline\lstinline
\DeclareRobustCommand*{\lstinline}{
  \ifmmode
    \let\SavedBGroup\bgroup
    \def\bgroup{
      \let\bgroup\SavedBGroup
      \hbox\bgroup
    }
  \fi
  \SavedLstInline
}

\AtBeginDocument{

}

\definecolor{orange}{RGB}{255,127,0}
\definecolor{grey}{RGB}{135,135,135}

\lstdefinelanguage{diff}{
  language=java,
  basicstyle=\ttfamily\scriptsize,
  sensitive=true,
  morecomment=[f][\color{gray}][0]{diff},
  morecomment=[f][\color{gray}][0]{index},
  morecomment=[f][\color{blue}][0]{@@},
  morecomment=[f][\color{magenta}][0]{***},
  morecomment=[f][\color{violet}][0]{!},
  morecomment=[f][\color{red!60!black}][0]{-},
  morecomment=[f][\color{green!60!black}][0]{+},
  morecomment=[f][\color{magenta}][0]{---},
  morecomment=[f][\color{magenta}][0]{+++},
  morecomment=[f][\color{gray}][0]{Binary},
  morecomment=[f][\color{gray}][0]{Only},
  morecomment=[f][\color{gray}][0]{old},
  morecomment=[f][\color{gray}][0]{new},
  morecomment=[f][\color{gray}][0]{rename},
  morecomment=[f][\color{gray}][0]{similarity},
  morecomment=[f][\color{gray}][0]{deleted},
  morecomment=[f][\color{magenta}][0]{***************},
  morecomment=[f][\color{red!60!black}][0]<,
  morecomment=[f][\color{green!60!black}][0]>,
  morecomment=[f][\color{blue}][0]{0},
  morecomment=[f][\color{blue}][0]{1},
  morecomment=[f][\color{blue}][0]{2},
  morecomment=[f][\color{blue}][0]{3},
  morecomment=[f][\color{blue}][0]{4},
  morecomment=[f][\color{blue}][0]{5},
  morecomment=[f][\color{blue}][0]{6},
  morecomment=[f][\color{blue}][0]{7},
  morecomment=[f][\color{blue}][0]{8},
  morecomment=[f][\color{blue}][0]{9},
}[comments]

\newcommand\nbStrategies{9\xspace}
\newcommand\nbBugsInBenchmark{16\xspace}

\newcommand{\mycode}[1]{{\small \texttt{#1}}\xspace}

\newcommand\NPEFix{NPEfix\xspace}

\newcommand\stratReplaceVarLocal{S1a\xspace}
\newcommand\stratReplaceVarGlobal{S1b\xspace}

\newcommand\stratReplaceNewLocal{S2a\xspace}
\newcommand\stratReplaceNewGlobal{S2b\xspace}
\newcommand\stratSkipLine{S3\xspace}
\newcommand\stratReturnNull{S4a\xspace}
\newcommand\stratReturnVar{S4c\xspace}
\newcommand\stratReturnNew{S4b\xspace}
\newcommand\stratReturnVoid{S4d\xspace}

\begin{document}

\title{Dynamic Patch Generation for Null Pointer Exceptions using Metaprogramming}
\author{Thomas Durieux, Benoit Cornu, Lionel Seinturier, and Martin Monperrus\\University of Lille \& Inria, France}
\maketitle

\label{chap:NPEFix}

\begin{abstract}
Null pointer exceptions (NPE) are the number one cause of uncaught crashing exceptions in production. In this paper, we aim at exploring the search space of possible patches for null pointer exceptions  with metaprogramming. 
Our idea is to transform the program under repair with automated code transformation, so as to obtain a metaprogram. 
This metaprogram contains automatically injected hooks, that can be activated to emulate a null pointer exception patch.
This enables us to perform a fine-grain analysis of the runtime context of null pointer exceptions.
We set up an experiment with \nbBugsInBenchmark real null pointer exceptions that have happened in the field. We compare the effectiveness of our metaprogramming approach against simple templates for repairing null pointer exceptions.
\end{abstract}

\section{Introduction}

Null pointer exceptions are the number one cause of uncaught crashing exceptions in production \cite{li2006have}.
Li et al. found that 37.2\% of all memory errors in Mozilla and Apache are due to null values~\cite{li2006have}.
Other studies\footnote{\url{http://bit.ly/2et3t79}, last accessed Oct 19 2016} found that up to 70\% of errors in Java production applications are due to null pointer exceptions.
It is an inherent fragility of software in programming languages where null pointers are allowed, such as C and Java. 
A single null pointer can make a request or a task to fail, and in the worst case, can crash an entire application.
Program repair \cite{Monperrus2015} is a research field concerned with the generation of patches.
There are a number of program repair techniques (eg. \cite{le2012genprog}) but not one of them is dedicated to null pointer exceptions.

One way of fixing null pointer exceptions is to use a template \cite{Kim2013}.
For instance, one can reuse an existing local variable as follows:
\begin{lstlisting}[language=diff]
+ if (r == null) {
+   r = anotherVar;
+ }
  r.foo(p);
\end{lstlisting}

This example illustrates a template that depends on the context (called parametrized template in this paper): the reused variable is the template parameter.
When repairing a statically typed language, such as Java that we consider in this paper, the static type of the template parameter has to be compatible with the variable responsible for the null pointer exception. This is done with static analysis of the code.

However, there may be other variables in the context of the null pointer exception, typed by a too generic class, which may be compatible. 
\emph{The static analysis of the context at the location of the null pointer exception may not give a complete picture of the objects and variables that can be used for fixing a null pointer exception.} There is a need to make a dynamic analysis of the repair context. This is the problem we address in this paper.

To overcome this problem, we propose to perform a dynamic analysis of the repair context. 
This completely changes the way we perform patch generation: instead of applying templates one after the other, we propose to create a metaprogram that is able to dynamically analyze the runtime context, and to dynamically alter the execution of the program under repair to identify a working patch.
The resulting technique is called \NPEFix.
It takes as input a test case that triggers an uncaught null pointer exception and outputs a patch that fixes it if one exists (or several alternative patches if several valid patches exist).
\NPEFix is composed of three main phases: first, it generates a metaprogram out of the program under repair, second it compiles the metaprogram, third, it runs the failing test case a large number of times, each time altering the behavior to find a way to avoid the null pointer exception.

\NPEFix is based on repair strategies for null pointer exceptions. 
Reusing an existing variable, as seen above, is one such strategy. 
\NPEFix uses a total of \nbStrategies different strategies that are categorized in two groups. 
The first group is about providing an alternative value when a null pointer exception is about to happen. 
This value can come from elsewhere in the memory (i.e. a valid value that is stored in another variable), or it can be a new object. 
The second group of strategies is about skipping the execution of the exception-raising expression.
It can be either skipping a single statement or skipping a complete method.

In \NPEFix, we use a metaprogramming approach -- a code transformation -- for each of those strategies.
All transformations are compatible one with another and can be used in conjunction.
Then, we explore the search space of tentative patches for null pointer exceptions by activating the hooks of the metaprogram which have been injected by transformation. 
Compared to a template based approach, \emph{the novelty of our approach is to explore the search space of tentative patches purely at run-time, which gives us a way to finely analyze the runtime context of the null pointer exception}. 

To evaluate our approach, we build a benchmark of \nbBugsInBenchmark null pointer exceptions that happened in the field for open-source software and were all reported in bug trackers. For each of those bugs, we create the \NPEFix metaprogram, and explore the search space to identify patches that are able to avoid the null pointer exception to happen.

To sum up, the contributions of this paper are:
\begin{itemize}
\item A taxonomy of \nbStrategies alternative execution strategies to repair null pointer exceptions (\autoref{sec:taxonomy}).
\item A set of code transformations for implementing those null-pointer repair strategies in Java (\autoref{sec:implementation}).
\item The systematic evaluation of our approach on \nbBugsInBenchmark real null pointer exceptions that we were able to reproduce (\autoref{sec:eval}). We compare \NPEFix against naive template-based repair.
\item A publicly available tool and benchmark for facilitating further research on this topic: \url{https://github.com/Spirals-Team/npefix}.
\end{itemize}

This paper is a reworked and extended version of a working paper published on Arxiv \cite{cornu:hal-01251960}.
It is structured as follows.
Section \ref{sec:taxonomy} presents a taxonomy of repair strategies for null pointer exceptions.
Section \ref{sec:implementation} describes our metaprogramming approach to apply those repair strategies.
Section \ref{sec:eval} discusses the evaluation of our work.
Section \ref{sec:discussion}, \ref{sec:rw} respectively explores the limitations of our approach and the related work.

\section{A Taxonomy of Repair Strategies for Null Pointer Exceptions}
\label{sec:taxonomy}

In this section, we present a taxonomy of run-time repair strategies for null pointer exceptions. It unifies previous work on that topic \cite{dobolyi2008changing,kent2008dynamic,LongSR14} and proposes new strategies.

When a harmful null pointer exception is going to happen, there are two main ways to avoid it.
First, one can replace the null value by a valid object.
Second, one can skip the problematic statement.
In both cases, no null pointer exception happens.
In this paper, we refine those two techniques in \nbStrategies different strategies to repair null pointer exceptions.
There are grouped along those two families: replacement of the null by an alternative object, and skipping the execution of the statement associated with the null pointer exception.

\begin{table}
\caption{\NPEFix' repair for null pointer exceptions.}
\label{tab:strategies}
\begin{tabularx}{\columnwidth}{|l|l|l|l|X|}
\hline
\multicolumn{3}{|c|}{Strategy} &  Id & Description \\ \hline
\multirow{6}{*}{\rotatebox{90}{Replace}} 
 & \multirow{4}{*}{reuse}    & local    & \stratReplaceVarLocal  & local reuse of an existing compatible object \\ \cline{3-5}
 &                           & global   & \stratReplaceVarGlobal & global reuse of an existing compatible object \\ \cline{2-5}
 & \multirow{2}{*}{creation} & local    & \stratReplaceNewLocal  & local creation of a new object \\ \cline{3-5}
 &                           & global   & \stratReplaceNewGlobal & global creation of a new object \\ \hline
\multirow{6}{*}{\rotatebox{90}{skipping}} 
 & \multicolumn{2}{c|}{line}            & \stratSkipLine         & skip statement \\ \cline{2-5}
 & \multirow{5}{*}{\rotatebox{90}{method}} 
                             & null     & \stratReturnNull       & return a null to caller \\ \cline{3-5}
 &                           & creation & \stratReturnNew        & return a new object to caller \\ \cline{3-5}
 &                           & reuse    & \stratReturnVar        & return an existing compatible object to caller \\ \cline{3-5}
 &                           &          & \stratReturnVoid       & return to caller (void method)\\ \cline{3-5}
\hline
\end{tabularx}
\end{table}

\subsection{Strategies Based on Null Replacement}
One way to avoid a null pointer exception to happen is to change the reference into a valid instance.
We can provide an existing value (if one can be found) or a new value (if one can be constructed).
To facilitate the presentation, \emph{r} is an expression (usually a variable reference).
We basically want \emph{r} to reference a valid (non-null) value in order to prevent a null pointer exception when executing r.foo(p).
Symbol $p$ is a method parameter. 

\textbf{Reuse (\stratReplaceVarGlobal)} A first strategy is the case of repairing the null pointer exception with an existing object as follows:
\begin{lstlisting}[language=diff]
+ if (r == null) {
+   r = anotherVar;
+ }
  r.foo(p);
\end{lstlisting}

Strategy \stratReplaceVarGlobal is parameterized by the variable $anotherVar$.
The variable is taken from the set $S$ of accessible objects composed of the local variables, the method parameters, the class fields of the current class and all the other static variables.
$S$ is further filtered to only select the set of all well typed and non-null values \emph{V} according to the type of $r$.

\textbf{Local Reuse (\stratReplaceVarLocal)} 
A variant of the reuse strategy consists in replacing one null reference by a valid object, without modifying the null variable itself.

\begin{lstlisting}[language=diff]
+ if (r == null) {
+   anotherVar.foo(p);
+ } else {
    r.foo(p);
+ }
\end{lstlisting}

With local reuse, all the other statements using \emph{r} will still perform their operations on \emph{null}.

\textbf{Object Creation (\stratReplaceNewGlobal)} 
Another strategy consists of creating a new value.
\begin{lstlisting}[language=diff]
+ if (r == null) {
+   r = new Foo();
+ }
  r.foo();
\end{lstlisting}

\textbf{Object Creation Local (\stratReplaceNewLocal)} 
A rare possible patch for a null pointer exception consists of providing a disposable object. This is what we call local object creation. This is interesting if method ``foo'' changes the state of p based on the method call receiver. 

\begin{lstlisting}[language=diff]
+ if (r == null) {
+   new Foo().foo(p);
+ } else {
    r.foo(p);
+ }
\end{lstlisting}

This sums up in 4 possible strategies for  null replacement (see \autoref{tab:strategies}):
use an existing value locally (\stratReplaceVarLocal), 
use an existing value globally (\stratReplaceVarGlobal), 
use a new value locally (\stratReplaceNewLocal) and use a new value globally (\stratReplaceNewGlobal).

\subsection{Strategies Based on Execution Skipping}
We also propose to skip the statement where a null pointer exception would happen.
There are different possible ways of skipping.

\textbf{Line Skipping (\stratSkipLine)} 
First, the straight-forward strategy \stratSkipLine consists in skipping only the problematic statement and allows to avoid the null pointer exception at this location.

\begin{lstlisting}[language=diff]
+ if (r != null) {
   r.foo(p);
+ }
\end{lstlisting}

We also propose a family of strategies which consists in skipping the rest of the method.
For skipping the rest of the method, several 
possibilities can be considered.
Let us consider that the method returns an object of type ``Bar``.

\textbf{Return Null \stratReturnNull} 
If the method expects a return value, one can return null: this is a reasonable option because it is possible that the caller has a non-null check on the returned object.

\begin{lstlisting}[language=diff]
+ if (r == null) {
+   return null; 
+ }
  r.foo(p);
\end{lstlisting}

\textbf{Return New Object \stratReturnNew} 
One can return a new instance of the expected type. As for strategy \stratReplaceNewGlobal, this is a strategy parameterized by all possible constructor calls. 
\begin{lstlisting}[language=diff]
+ if (r == null) {
+   return new Bar(); 
+ }
  r.foo(p);
\end{lstlisting}

\textbf{Return Variable \stratReturnVar} 
One can also search in the set of  accessible values one which corresponds to the expected return type and return it. As for strategy \stratReplaceVarGlobal, this is a strategy parametrized by all possible type-compatible variables. 
\begin{lstlisting}[language=diff]
+ if (r == null) {
+   return anotherVar; 
+ }
  r.foo(p);
\end{lstlisting}

\textbf{Vanille Return \stratReturnVoid} When the method does not return anything (\texttt{void} return type in Java), inserting a simple ``\verb|if (r==null) { return; }|'' is a valid option.

All strategies are listed in \autoref{tab:strategies}.
The table represents the different dimensions of the analysis:
replacement vs skipping, local vs global, reusing objects vs creating new ones.
For each strategy, the corresponding code that needs to be injected is shown in the last column.

\subsection{Novelty}

Among those \nbStrategies strategies, some of them have already been explored.
Dobolyi et al. \cite{dobolyi2008changing} have proposed two of them: \stratReplaceNewGlobal and \stratSkipLine.
Kent \cite{kent2008dynamic} have defined \stratReplaceNewGlobal, \stratSkipLine, \stratReturnNull and \stratReturnVoid.
Long et al. \cite{LongSR14} have explored \stratSkipLine, \stratReturnNew and \stratReturnVoid.
This means that $4/\nbStrategies$ strategies presented in this paper are new:
\stratReplaceVarLocal,
\stratReplaceVarGlobal,
\stratReplaceNewLocal,
\stratReturnVar.
The novelty lies on the idea of reusing existing objects (\stratReplaceVarLocal, \stratReplaceVarGlobal, \stratReturnVar) and performing local injection (\stratReplaceVarLocal, \stratReplaceNewLocal).

\subsection{A Naive Implementation for NPE Repair}
\label{sec:naive-implementation}

To find patches according to the strategies presented in \autoref{sec:taxonomy}, a naive implementation consists of exploring all possible strategies one by one. 
We call this naive implementation TemplateNPE, whose main algorithm is shown in Algorithm \ref{algo:naive}. The idea of this algorithm is to explore one strategy after each strategy, using source code transformation techniques, and to test whether they repair the null pointer exception under consideration.
If the application of a template compiles, we have a \emph{``tentative patch''}.
Hence, this algorithm explores the search space of all tentative patches.
One sees that this naive implementation requires recompiling one file for each pair $(parameter, strategy)$.

\begin{figure}[t]
  \begin{algorithmic}[1]
    
    \REQUIRE{p: a program}
    \REQUIRE{t: a test case reproducing a NPE}
    \REQUIRE{e: expression that triggers the NPE}
    \REQUIRE{S: a set of repair strategies}
    \ENSURE{P: set of tentative patches}
    \ENSURE{Q: set of valid patches}

    \STATE{compile $p$}
    \FOR{s in S}
        \STATE{A $\gets$  possible parameter value for $s$ in $e$}
        \FOR{$a$ : A}
            \STATE{$x$ $\gets$  apply $s$ on $p$ parametrized by $a$}
            \STATE{recompile $x$}
            \IF{x compiles}
                \STATE{add $x$ to P}
            \ENDIF
            \STATE{run $t$ against $x$}
            \IF{$t$ succeeds}
                \STATE{add $x$ to Q}
            \ENDIF
        \ENDFOR 
    \ENDFOR
  \end{algorithmic}
  \caption{TemplateNPE: Exploration of all tentative patches based on parametrized templates}
  \label{algo:naive}
\end{figure}

All patch templates are parametrized.
The first template parameter contains the Java expression that may be null, it is located in the condition of the template as follows: \verb|if (<parameter> == null) ... |.
The second template parameter refers to the expression that is used to replace the null expression.
This expression can be a variable (\stratReplaceVarLocal, \stratReplaceNewGlobal), a new instance  or a predefined constant (null, 0, 1, "", ' ') for \stratReplaceNewLocal, \stratReplaceNewGlobal.
This template parameter is not present in \stratSkipLine, \stratReturnNull and \stratSkipLine).
The replacement expressions are statically created  based on the static analysis of the context of the line that produces the null dereference.
We use code analysis to perform which variables are accessible at the expression causing the null pointer exception, and also to list constructor calls for building new instances.

\section{Metaprogramming for Null Pointer Repair}
\label{sec:implementation}

We now present a set of code transformations to embed the \nbStrategies strategies presented in \autoref{sec:taxonomy} in a metaprogram. 

In this paper, a metaprogram is a program enriched with behavior modification hooks.
By default, all behavior modification hooks are deactivated, which means that by default a metaprogram is semantically equivalent to the original program. 

Let us consider the program \mycode{x = y + z}. A metaprogram is for instance \mycode{x = y - z if HOOK\_IS\_ACTIVATED else y + z} (functional style) or \mycode{x = HOOK\_IS\_ACTIVATED ? y - z : y + z} (ternary expression, e.g. in Java).
Variable HOOK\_IS\_ACTIVATED is a Boolean global variable which controls the activation of the behavior modification. 
This metaprogram enables one to transform at run-time an addition into a subtraction. In our context, the metaprogram is automatically created using source-to-source transformation.

The metaprogramming code transformations are realized in a tool for Java called \NPEFix.
\NPEFix is composed of three main phases: first, it generates a metaprogram out of the program under repair, second it compiles the metaprogram, third, it runs the failing test case  a large number of times, each time altering the behavior to find a way to avoid the null pointer exception and thus emulate a patch.

\subsection{Core Intercession Hook}

To modify the behavior when a null pointer exception happens, we encapsulate each method call and field access as shown in \autoref{fig:npe-encaps}.

The call of \mycode{doSomething} that is originally present is now made on the result of method \mycode{check\-For\-Null}.
Method \mycode{check\-For\-Null} does the following things.
It first assesses whether the object is null, i.e. whether a null pointer exception will occur; if it's not null, the program proceeds with its normal execution.
If the object is null, it means a null pointer exception is about to be thrown, and then a strategy is applied. 
For sake of simplification, this is shown as a switch as in  Listing \ref{fig:npe-encaps}, this switch case is the core intercession hook of the metaprogram.

\begin{lstlisting}[numbers=none,caption={Detecting harmful null pointer exceptions With Code Transformation},label=fig:npe-encaps,float]
//before transformation
Foo o;
o.doSomething();

// after NPEfix transformation
checkForNull(o, Foo.class).doSomething();

// with static method checkForNull
Object checkForNull(Object o, Class c){
  if (o == null) // null pointer exception detected
    switch (STRATEGY) {
      case s1b: return getVar(currentMethod());
      case s2b: return createObject(c);       
      ...
    }
  return o;
}
\end{lstlisting}

\subsection{Value Replacement Strategies}
There are four strategies based on value replacement (the first half of \autoref{tab:strategies}):
\stratReplaceVarLocal, \stratReplaceVarGlobal, \stratReplaceNewLocal and \stratReplaceNewGlobal.

\subsubsection{Reuse Variable}

\begin{lstlisting}[numbers=none,caption={Maintaining a set of variables as pool for replacement at run-time}, label=fig:npe-getVar,float]
//before transformation
public void method(){
    ...
    Object a = {defaultExpression};
    a = {newValue};
    ...
}

// after NPEfix transformation
public void method(){
    collectField(myField, "myField");
    ...
    Object a = initVar({defaultExpression}, "a");
    a = modifyVar({newValue}, "a");
    ...
}
\end{lstlisting}

For replacing a null value with a variable, the challenge is to maintain a set of variables as a pool for replacement at run-time.
\autoref{fig:npe-getVar} shows how we tackle this problem; we use a stack to store all the variables of each method.
Each variable initialization and assignment inside the method is registered thanks to the  \NPEFix' method \mycode{initVar}.
In addition, at the beginning of each method, we collect all the accessible fields and parameters.

\subsubsection{Create New Object}
Now, let us consider that the strategies that create a new variable (strategies \stratReplaceNewLocal and \stratReplaceNewGlobal).

As shown in \autoref{fig:npe-encaps}, a call is made to \mycode{createObject} that takes as parameter the static type of the dereferenced variable.
\mycode{createObject} uses reflection to access to all the constructors of the given type.
In addition, this method is recursive so as to create complex objects if needed. 
It tries to create a new instance of the class from each available constructor.
Given a constructor, it attempts to create a new instance for each of the parameter recursively.
The stopping condition is when a constructor does not need parameters.
Note that the primitive types, which don't have constructors, are also handled with default literal values.

\subsection{Skipping Strategies}

Now we present how we implement the strategies based on skipping the execution (the second half of \autoref{tab:strategies}).

\subsubsection{Line skipping}

\begin{lstlisting}[numbers=none,caption={Implementation of Line-based Skipping},label=fig:npe-skipLine,float]
// before transformation
value.dereference(); 

// after NPEfix transformation
if (skipLine(value)){ 
    value.dereference();
}
boolean skipLine(Object... objs){ // NPEfix framework
    for (Object o : objs) {
        if (o == null && cannotCatchNPE() && doSkip())
            return false
    }
    return true;
}
\end{lstlisting}

The strategy \stratSkipLine necessitates to know if a null pointer exception will happen in a line, before the execution of the line.
For this, the transformation presented in \autoref{fig:npe-encaps} is not sufficient, because the call to method  \mycode{checkForNull} implies that the execution of the line has already started.
To overcome this issue, we employ an additional transformation presented in \autoref{fig:npe-skipLine}.

Similarly to \mycode{checkForNull}, method \mycode{skipLine} assesses, before the line execution, whether the dereferenced value is null or not, and whether it is harmful.
Method skipline takes an arbitrary number of objects, the ones that are dereferenced in the statement. This list is extracted statically.

\subsubsection{Method skipping}
The remaining strategies are based on skipping the execution of the rest of the method when a harmful dereference is about to happen: these are strategies \stratReturnVoid, \stratReturnNull, \stratReturnVar and \stratReturnNew (the last part of \autoref{tab:strategies}).
We implement those strategies with a code transformation as follows.

A try-catch block is added in all methods, wrapping the complete method body.
This try-catch blocks handle a particular type of exception defined in our framework (ForceReturnError).
This exception is thrown by the \mycode{skipLine} method when one of the method-skipping strategies is activated, as show in \autoref{fig:npe-skipMeth}.
This listing also shows a minimalist example of the code resulting from this transformation.

\begin{lstlisting}[numbers=none,caption={Metaprogramming for method-based skipping strategies},label=fig:npe-skipMeth,float]
// before transformation
Object method(){
  ...
  value.dereference();
  ...
  return X;
}
// after NPEfix transformation
Object method(){
  try {
    ...
    if (skipLine(value)){
      value.dereference();
    }
    ...
    return X;
  } catch (ForceReturnError f){
    if (s4a) return null;
    if (s4b) return getVar(Object.class);
    if (s4c) return createObject(Object.class);
  } 
}
boolean skipLine(Object... objs){
  if(hasNull(objs) && cannotCatchNPE() && skipMethodActivated())
    throw new ForceReturnError();
  ...
}
\end{lstlisting}

\subsection{Exploration of the Patch Search Space at Runtime}

Now that we have a metaprogram that embeds all strategies, we also have a way to explore the search space of null pointer repair purely at run-time. The idea is to first create the metaprogram, then to activate the strategies dynamically by setting the appropriate behavior modification hooks. 

\vspace{2mm}
\begin{algorithm}[t]
  \begin{algorithmic}[1]
    \REQUIRE{p: a program}
    \REQUIRE{t: a test case reproducing a NPE}
    \REQUIRE{S: a set of repair strategies}
    \ENSURE{P: set of tentative patches}
    \ENSURE{Q: set of valid patches}
    \STATE{$M_p$ $\gets$ create meta-program of p}
    \STATE{compile $M_p$}

    \STATE{execute $t$ against $M_p$ until null dereference}
    \STATE{D $\gets$  collects all possible metaprogramming patches based on runtime analysis}
    \STATE{terminate execution}
    \FOR{d in D}
        \STATE{activate d in metaprogram}
        \STATE{execute $t$ against $M_p$}
        \STATE{p$\gets$  patch corresponding to $d$}
        \STATE{add $p$ to P}
        
        \IF{t succeeds}
            \STATE{add $p$ to Q}
        \ENDIF

    \ENDFOR
  \end{algorithmic}
  \caption{The Exploration Algorithm of \NPEFix}
  \label{algo:exploration}
\end{algorithm}

\textbf{From Hooks to Patches} Given a combination of behavioral modification hooks, one can create the corresponding source code patch by reinterpreting the hooks according to the templates presented in \autoref{sec:taxonomy}.

\subsection{Implementation}
All those transformations have been implemented in a tool called \NPEFix, which has been made publicly available for sake of reproducible research and open science\footnote{\url{https://github.com/Spirals-Team/npefix}}. 
The transformations use the Spoon library \cite{spoon}.

\section{Evaluation}
\label{sec:eval}

We now evaluate \NPEFix. 
We design a protocol to answer the following questions.

\newcommand{\rqSpace}{RQ1. What is the impact of \NPEFix' runtime analysis of the repair context and TemplateNPE's static analysis of the repair context on the number of explored tentative patches?}
\newcommand{\rqTSAdequatePatches}{RQ2. Does \NPEFix (metaprogramming) produce more valid patches?}
\newcommand{\rqCaseStudies}{RQ3. What are the reasons explaining the presence of different valid patches?}
\newcommand{\rqPerformance}{RQ4. Is the performance of  \NPEFix acceptable?}

\begin{itemize}
\item \rqSpace
\item \rqTSAdequatePatches
\item \rqCaseStudies
\item \rqPerformance
\end{itemize}

\subsection{Protocol}

In order to evaluate our approach of patch generation based on metaprogramming, we build a benchmark of real and reproducible null pointer exceptions in Java programs (see \autoref{sec:dataset}).
Then we compare the ability of TemplateNPE and \NPEFix  to find different patches and repair each bug of the benchmark.
TemplateNPE is the template-based implementation of our nine repair strategies that is described in \autoref{sec:naive-implementation}.
\NPEFix uses our metaprogramming approach to fix null pointer exceptions as described in \autoref{sec:implementation}. 

In this paper, we study ``test-suite adequate'' patches \cite{martinez2016}
We consider a patch as test-suite adequate when the failing test case that reproduces the null dereference passes as follows:
1) no exception (a null pointer exception or another one) is uncaught and crashes the test case and
2) whether the assertions that come at the end of the test case reproducing the null pointer exception  pass.
For sake of readability, we refer to test-suite adequate patches as ``valid'' patches
(hence we use test-suite adequate and ``valid'' interchangeably in this paper)

Note that test-suite adequate patches mare be conceptually incorrect, even if they pass the test suite.
Since we use as only oracle the test cases, which can miss correctness assertions, a valid patch can be considered incorrect beyond the test-suite satisfaction correctness criterion \cite{Long2016analysis,nopol}. 

\subsection{Evaluation Metrics}

After each bug and each repair technique, we collect three metrics:
the number of tentative patches (whether valid or not);
the number of valid patches (that pass the test case);
the execution time required to explore the search space of patches.

We interpret those metrics as follows.
A larger number of tentative patches means that the search space is richer. 
As shown in previous work \cite{martinez2016,Long2016analysis}, there are often multiple different patches that are able to make a failing test case passing.
A larger number of generated valid patches is better, it means that the developer is given more choices to get a really good choice.

The results of this experimentation, incl. all tentative patches are publicly available at
\url{https://github.com/Spirals-Team/npefix-experiments}

\subsection{Benchmark}\label{sec:dataset}

\begin{table}
\caption{Descriptive statistics of the subjects suffering from null pointer exceptions. Size measured by cloc version 1.60.}
\label{tab:desc-stats}
\begin{tabularx}{\columnwidth}{|l|X|r|}
\hline
Subject &Domain           & Size        \\
\hline
COLL & Collection library & \numprint{21594} LOC\\
LANG & Utility functions & \numprint{18970} LOC\\
MATH & Math library & \numprint{90771} LOC\\
PDF & PDFBox library & \numprint{64375} LOC\\
Felix & Felix library & \numprint{33057} LOC\\
SLING & Sling library & \numprint{583} LOC\\
\hline
Total & \nbBugsInBenchmark bugs & \numprint{229350} LOC\\
\hline
\end{tabularx}
\end{table}

To build a benchmark of real null pointer exceptions in Java, we consider two inclusion criteria.
First, the bug must be a real bug reported on a publicly available forum (e.g. a bug tracker).
Second, the bug must be reproducible.
This point is very challenging since it is really difficult to reproduce field failures, due to the absence of the exact crashing input, or the exact configuration information (versions of dependencies, execution environment, etc.). As a rule of thumb, it takes one day to find and reproduce a single null pointer exception bug.
We consider bugs in the Apache Commons set of libraries (e.g. Apache Commons Lang) because they are well-known, vastly used and their bug trackers are public, easy to access and to be searched.
Also, thanks to the strong software engineering discipline of the Apache foundation, a failing test case is often provided in the bug report.
We have not rejected a single reproducible field null pointer exception.

As a result, we have a benchmark that contains \nbBugsInBenchmark null pointer exceptions (1 from collections, 3 from lang and 7 from math, 3 from PDFBox, 1 from Felix, 1 from Sling). 
It is publicly available for future research (\url{https://github.com/Spirals-Team/npe-dataset}).
The main strength of this benchmark is that it only contains real null pointer exception bugs and no artificial or toy bugs.
\autoref{tab:desc-stats} shows the size of applications for which we have real field failures.

\subsection{\rqSpace}
\label{sec:rq1}

\begin{table*}[t]
\centering
\caption{Comparison of the the template approach (TemplateNPE) and the meta-programming approach (\NPEFix) on three key metrics.}
\label{tab:repair_space}
\setlength\tabcolsep{0.7 ex}
\begin{tabularx}{0.7\textwidth}{|X|r|r||r|r||r|r|}\hline
\multirow{3}{*}{Bug ID} & \multicolumn{2}{c||}{\# Tentative Patches} & \multicolumn{2}{c||}{\# Valid Patches}  & \multicolumn{2}{c|}{Execution Time}  \\\cline{2-7}
                 & Template  & \NPEFix   &    Template  &    \NPEFix   &    Template  &    \NPEFix  \\\cline{2-7}
 \hline
 collections360  &         7 & \textbf{10} &            0 &            0 & 0:00:20      & 0:02:32     \\
 felix-4960      &         9 &         7 &            4 &            4 & 0:00:43      & 0:03:22     \\
 lang304         &        44 &\textbf{77} &           43 &  \textbf{65} & 0:00:11      & 0:00:26     \\
 lang587         &        21 &\textbf{28} &           12 &  \textbf{28} & 0:00:23      & 0:00:23     \\
 lang703         &        16 &        15 &            0 &    \textbf{7} & 0:00:09      & 0:00:20     \\
 math1115        &         8 &\textbf{11} &            6 &            5 & 0:00:47      & 0:02:08     \\
 math1117        &         8 &\textbf{11} &            0 &            0 & 0:00:41      & 0:01:52     \\
 math290         &         9 &\textbf{10} &            3 &    \textbf{4} & 0:00:18      & 0:00:42     \\
 math305         &         2 &\textbf{4} &            1 &     \textbf{3} & 0:00:08      & 0:00:40     \\
 math369         &        15 &\textbf{16} &           14 &           14 & 0:00:26      & 0:00:43     \\
 math988a        &        17 &        17 &           11 &           11 & 0:01:02      & 0:01:38     \\
 math988b        &        17 &\textbf{25} &           17 &           17 & 0:01:39      & 0:01:48     \\
 pdfbox-2812     &         6 &\textbf{14} &            2 &            2 & 0:00:28      & 0:01:46     \\
 pdfbox-2965     &         4 &         3 &            4 &            3 & 0:00:16      & 0:01:34     \\
 pdfbox-2995     &         4 &\textbf{5} &            3 &            1 & 0:00:13      & 0:01:35     \\
 sling-4982      &        18 &\textbf{20} &            7 &   \textbf{11} & 0:00:05      & 0:00:06     \\
\hline
 Total           &       205 &       273 &          127 &          175 & 0:07:57      & 0:21:44     \\
 Average         & 12.81     & 17.06     & 7.94         & 10.94        & 0:00:29      & 0:01:21     \\
 Median          & 9.00      & 12.50     & 4.00         & 4.50         & 0:00:22      & 0:01:35     \\ \hline

\end{tabularx}
\end{table*}

\autoref{tab:repair_space} presents the results of our experiment.
The first column contains the bug identifier.
The second column contains the number of tentative patches for each bug (ie. the size of the search space).
This column is composed of two sub-columns: the number of tentative patches using the template-based approach (TemplateNPE) and the number of patches generated by the metaprogramming approach (\NPEFix).
For example, TemplateNPE identifies 7 tentative patches and \NPEFix identifies 10 tentative patches.
The remaining top-level columns will discussed below in \autoref{sec:rq2} and \autoref{sec:rq3}.

In 12/\nbBugsInBenchmark (in bold) of the case \NPEFix explores more tentative patches than TemplateNPE.
This validates that the static analysis and dynamic analysis of the repair context differs, and that the latter is potentially richer.
The difference in the number of tentative patches between TemplateNPE and \NPEFix is explained as follows.
\begin{itemize}[leftmargin=*]

\item During execution, more objects are detected as compatible with the type of the null expression.
With the template approach, we do not know the actual runtime type of all variables.

\item The number of different new objects created varies because \NPEFix detects at runtime more compatible constructors.

\item Some strategies cannot be applied at certain locations with the template approach. For example, TemplateNPE cannot apply the skip-line strategy (\stratSkipLine) on a local variable. 
This  case is naturally handled in the metaprogramming approach.

\item In 3/\nbBugsInBenchmark cases, the template-based approaches identifies more tentative patches. The reason is that \NPEFix filters out equivalent patches by on the runtime variable value. This is further discussed in \autoref{case-study-Felix-4960} and \autoref{case-study-PDFBbox-2965}.
\end{itemize}

\begin{framed}
\textbf{\rqSpace}
\NPEFix explores 273 tentative patches, which is 68 more than the template-based approach.
In other words, the search space of the metaprogramming technique is larger. 
This validates our intuition that the runtime analysis of the repair context is valuable in certain cases.
\end{framed}

\subsection{\rqTSAdequatePatches}
\label{sec:rq2}

\autoref{tab:repair_space} also gives the number of tentative patches that are valid, ie. that make the failing test case passing. 
This is is shown in the 4th and 5th columns under the top-level header ``\# Valid Patches''
presents the results of this experiment.
For example, for bug Collection-360, neither TemplateNPE nor \NPEFix identifies a valid patch.
For lang304, TemplateNPE identifies 43 valid patches, while \NPEFix finds 65 valid patches.
As we can see there is a correlation between the size of the explored search space and number of valid patches identified. 
This means that it is worth exploring more tentative patches to identify more valid patches.

\begin{framed}
\textbf{\rqTSAdequatePatches}
\NPEFix finds 175 patches that avoid the null pointer exception and make the test case passing. Among those valid patches, 48 patches of them are uniquely found by the metaprogramming approach thanks to the runtime analysis of the repair context.
Those patches are only test-suite adequate, and if the test suite is weak, they may be incorrect.
\end{framed}

\subsection{\rqCaseStudies}
We answer to this research question with 3  case studies.

\subsubsection{Math-305}
\label{case-study-Math-305}

\begin{figure}
\begin{lstlisting}[language=diff,rulecolor=\color{green!60!black}]
--- org/apache/commons/math/stat/clustering/KMeansPlusPlusClusterer.java
+++ org/apache/commons/math/stat/clustering/KMeansPlusPlusClusterer.java
@@ -90,3 +90,7 @@
             Cluster<T> cluster = getNearestCluster(clusters, p);
-            cluster.addPoint(p);
+            if (cluster == null) {
+                     new Cluster(null).addPoint(p);
+            } else {
+                    cluster.addPoint(p);
+            }
         }
\end{lstlisting}
\begin{lstlisting}[language=diff,rulecolor=\color{green!60!black}]
--- org/apache/commons/math/stat/clustering/KMeansPlusPlusClusterer.java
+++ org/apache/commons/math/stat/clustering/KMeansPlusPlusClusterer.java
@@ -90,2 +90,5 @@
             Cluster<T> cluster = getNearestCluster(clusters, p);
+            if (cluster == null) {
+                    cluster = new Cluster(null);
+            }
             cluster.addPoint(p);
\end{lstlisting}
\begin{lstlisting}[language=diff,rulecolor=\color{green!60!black}]
--- org/apache/commons/math/stat/clustering/KMeansPlusPlusClusterer.java
+++ org/apache/commons/math/stat/clustering/KMeansPlusPlusClusterer.java
@@ -90,3 +90,5 @@
             Cluster<T> cluster = getNearestCluster(clusters, p);
-            cluster.addPoint(p);
+            if (cluster != null) {
+                    cluster.addPoint(p);
+            }
         }
\end{lstlisting}
\begin{lstlisting}[language=diff,rulecolor=\color{red}]      
--- org/apache/commons/math/stat/clustering/KMeansPlusPlusClusterer.java
+++ org/apache/commons/math/stat/clustering/KMeansPlusPlusClusterer.java
@@ -90,2 +90,5 @@
             Cluster<T> cluster = getNearestCluster(clusters, p);
+            if (cluster == null) {
+                    return null;
+            }
             cluster.addPoint(p);
\end{lstlisting}
\caption{The generated patches for the bug Math-305, the patches that have a green border are valid patches and the patch that has a red border is the invalid patch.}
\label{fig:patches-305}
\end{figure}

We now discuss the patches found for bug Math-306, where the null pointer exception is thrown during the computation of a clustering algorithm called  Kmeans.
The null pointer exception is triggered when a point is added to the nearest cluster.
When the library first computes the nearest cluster, it fails because the current point is at a distance largest than Integer.MAXVALUE.
Then, the nearest cluster is set to null and a null pointer exception is thrown when the library tries to add the current point to it.

\NPEFix succeeds to identify 4 different patches for this bug. They are all presented in \autoref{fig:patches-305}.
The top three first patches (in green) of the figure are test-suite adequate: they all avoid the null pointer exception to be thrown.
As we see, they have the same behaviour: they skip the line that produces the null pointer exception. 
The first and the third patches create a new cluster that is never used in the application: this is useless but it works. 
The second patch skips the line that produces the null pointer exception, resulting in the point not being added to the cluster.
The last patch is invalid because it produces a division per zero later in the execution: the test case still fails (not with the original null pointer exception but with a division-by-zero exception).

This case study illustrates the fact many different valid patches indeed exist. However, while they are all equivalent according to the test case oracle, they are not equivalent for the developers: a developer would likely discard a patch that creates a temporary object only to avoid a null pointer exception.

\subsubsection{Felix-4960}
\label{case-study-Felix-4960}
Felix is an implementation of the OSGI component model.
Real bug Felix-4960 is about a 
null pointer exception that is thrown in  method ``getResourcesLocal(name)'' which is dedicated to  searching a resource in a path.
The null pointer exception appears when Felix does not succeed to get a list of resources in the  path.
In this case Felix tries to iterate on a list that is null, which triggers the null pointer exception.

For Felix-4960, 
TemplateNPE is surprisingly able to generate more tentative patches: \NPEFix generates 7 tentative patches and TemplateNPE generates 9 tentative patches that compile.
The reason is that template generates tentative patches based on reusing variables, however those patches are meaning less.
TemplateNPE fills the template parameters with  \mycode{m\_activationIncludes} and \mycode{m\_declaredNativeLibs}.
However, those variables are null because not initialized at this location in the code.
In other words, TemplateNPE generates a patch that replaces a null by a null, which obviously results in the same null pointer exception as before.

On the contrary, \NPEFix works at runtime, and hence knows that the actual value of  \mycode{m\_activationIncludes} and \mycode{m\_declaredNativeLibs} are null. Hence, it does not even tries them, because it knows in advance that such a tentative patch would be invalid.   
In this case, the runtime analysis of the context is interesting to discard incorrect patches early in the process.

\subsubsection{PDFBbox-2965} 
\label{case-study-PDFBbox-2965}

PDFBbox is a PDF rendering library.
This library allows one to read the properties of a PDF file.
PDFBbox-2965 happens when PDFBox searches for a specific field on a PDF that contains no PDF form.
Internally PDFBox iterates on all form fields of the PDF and compare the name of the field against the searched field.
But when the PDF contains no form, the list of fields is null, and a null pointer exception is thus triggered.

\NPEFix generates 3 tentative patches and TemplateNPE generates tentative  4 patches for this bug. 
They are all valid according to the test case.
As we see, TemplateNPE generates one additional valid patches.
This patch is shown in \autoref{pdfbox2965_addpatch}.
It consists of returning variable \mycode{retval} if the \mycode{fields} variable is null.
This patch is of low-quality because \mycode{retval} is always null in this case.
In other words, a simple \mycode{return null;} is a more explicit patch, and would better help the developer.
This explains why \NPEFix does not generate this patch, as for  Felix-4960, thanks to runtime analysis, \NPEFix knows that  \mycode{retval} is null, and that \mycode{return retval;} is equivalent to \mycode{return null;} (strategy \stratReturnNull).

\begin{lstlisting}[language=diff,rulecolor=\color{green!60!black},label=pdfbox2965_addpatch,caption=The additional patch of TemplateNPE for the bug PDFBbox-2965,float]
--- pdfbox/pdmodel/interactive/form/PDAcroForm.java
+++ pdfbox/pdmodel/interactive/form/PDAcroForm.java
@@ -250,2 +250,5 @@
 
+   if (fields == null) {
+       return retval; // reval is null
+   }
    for (int i = 0; i < fields.size() && retval == null; i++)
\end{lstlisting}

\subsection{\rqPerformance}
\label{sec:rq3}

The last column of \autoref{tab:repair_space} presents the execution time required to explore the whole search space of tentative patches for fixing the null pointer exception.
For example, for the bug Collections-360, 
TemplateNPE explores the search space in 20 seconds while \NPEFix requires 2 minutes 32 seconds. 

In most cases, the template based approach is faster. The reason is that creating the metaprogram takes a lot of time due to the complexity of code transformations. Also, the additional code injected in the metaprogram slows down each repair attempt, ie slows down each execution of the failing test case.
For \NPEFix, the complete exploration of the search space takes at most 3 minutes 22 seconds.
We consider this acceptable since a developer can wait for 3 minutes before offered a set of automatically tentative patches. 

\begin{framed}
\textbf{\rqPerformance}
The complete exploration of the search space of tentative patches is faster with TemplateNPE. 
This highlights a trade-off between the number of patches explored found and the time to wait.
\end{framed}

\section{Discussion}
\label{sec:discussion}

\subsection{Patch Readability and Templates}

One concern of automatic patch generation is the readability of the generated patches and its impact on the maintainability of applications \cite{fry2012human}.
Let us discuss the readability of \NPEFix patches.
\NPEFix patches repair one specific type of bug: null dereference.
We observe that most developers fix these bugs by adding a null check (\mycode{if(... != null)  something()}) before the null expression.
Most \NPEFix strategies resemble those human-written patches and thus and can be as easily understood and maintained as human patches.
Our experience with \NPEFix suggests that template-inspired patches enables one to encode -- if not enforce -- readable patches.

\subsection{Genetic Improvement and Metaprogramming}
Genetic improvement \cite{langdon2016genetic} refers to techniques that change the behavior of a program in order to improve a specific metric, for example the execution time.
There is an interesting relation between genetic improvement and \NPEFix.
Both are based on dynamically changing the behavior of an application.
In this perspective, making failing tests pass can be considered as a functional metric to be optimized. 
Indeed, we think that meta-programming techniques similar to that proposed in this paper could be used for genetic improvement by creating a space of different program behaviours to  explore.

\subsection{Threat to Validity}

A bug in the implementation of TemplateNPE or \NPEFix is the threat to the validity of the findings based on quantitative comparison presented in \autoref{tab:repair_space}.

Our benchmark is only composed of real bugs due to null pointer exceptions which is a strength. However, since they are all in Java and from 6 projects, there is a threat to the external validity of our findings. 
Other bugs and an implementation of TemplateNPE and \NPEFix in another language may uncover other behavioral differences.  

\subsection{Limitations}

We now summarize the main limitations of \NPEFix that we identified during our experiment.
First, \NPEFix is complex. As we have seen, the template-based approach is quite straightforward, while the metaprogramming approach is complex to debug and maintain. Due to the richness and complexity of Java, there may be cases where the metaprogramming approach slightly changes the behaviour of the application.
Second, as we have seen in \autoref{sec:rq3}, \NPEFix is slower than the template based approach. 
Third, as for all test-suite based patch generation technique, there is a risk of generating test-suite adequate, yet incorrect patches as suggested by  the Math-305 example.

\subsection{Reflections on TemplateNPE}

We would like to emphasize the fact TemplateNPE works surprisingly well. Initially, TemplateNPE was built as baseline comparison for \NPEFix. However, over time, we realized that it is much simpler to implement and maintain, while keeping comparatively good results, and while being faster.
This is an important lesson learned for us and for the metaprogramming research community: advanced code transformations and behavior modifications at runtime is not necessarily the best option. Simpler may be better.

\section{Related Work}
\label{sec:rw}

\subsection{Patch Generation}

The literature on patch generation is growing very fast. We only present a brief overview of notable contributions and refer to \cite{Monperrus2015} for a comprehensive overview.
GenProg by \cite{le2012genprog} applies genetic programming to the AST of a buggy program and generates patches by adding, deleting, or replacing AST nodes. 
Debroy and Wong \cite{DBLP:conf/icst/DebroyW10} propose a mutation-based repair method inspired from mutation testing. This work combines fault localization with program mutation to exhaustively explore a space of tentative patches.
SemFix by \cite{semfix} is a constraint based repair approach. It provides patches for assignments and conditions by combining symbolic execution and code synthesis. Nopol by  \cite{demarco2014automatic,nopol} is also a constraint based method, which focuses on repairing bugs in if-conditions and missing preconditions. SPR \cite{Long15} defines a set of staged repair operators so as to early discard  many candidate repairs that cannot pass the supplied test suite and eventually to exhaustively explore a small and valuable search space.

We have discussed in this paper a template-based patch generation approach.
PAR by \cite{Kim2013} uses 10 patch templates for common programming errors, Relifix \cite{reliflix} defines templates specifically for regression bugs. 
All of those approaches require a regression test suite to validate the patch, none of them leverage production traffic to assess the absence of regressions.

\subsection{Metaprogramming for Repair}

Rinard et al. \cite{rinard2004enhancing} presented a technique called ``failure oblivious computing'' to avoid illegal memory accesses.
Their idea is to create a metaprogram that adds additional code around each memory operation during the compilation process.
For example, the additional code verifies at run-time whether an array is accessed out of his bounds.
If the access is outside the allocated memory, the access is ignored (akin line skipping presented in this paper) instead of crashing with a segmentation fault.

Dobolyi and Weimer~\cite{dobolyi2008changing} present a technique to tolerate null pointer exceptions using metaprogramming.
Using code transformation, they introduce hooks to a recovery framework.
This framework is responsible for forward recovery of the form of creating a default object of an appropriate type of skipping instructions. Their strategies are a small subset of ours, and they do not explore the search space as we do in this paper.

The closest related work is by Kern and Esparza \cite{kern2010automatic} use a metaprogram that integrates all possible mutations according to a mutation operator. The mutations that are actually executed are driven by meta-variables.  A repair is a set of values for those meta-variables. The meta-variables are valued using symbolic execution. Both the metaprogram and the kind of faults are completely different from what we have presented in this paper.

\section{Conclusion}
\label{sec:conclusion}

In this paper, we have presented a novel and original technique to explore the repair search space of null pointer exception bugs.
Our technique, called \NPEFix, is based on \nbStrategies strategies that are specific to null pointer exceptions, and uses metaprogramming to explore the search space of all possible strategies.
We have evaluated our technique on \nbBugsInBenchmark real null pointer exceptions:
\NPEFix is able to successfully explore the search space of null pointer repair, and finds 175 valid patches over all considered bugs.

Future work is required to see whether other patch templates can be implemented using metaprogramming.
For instance, we are confident that even a generic technique as Genprog \cite{le2012genprog} could be implemented in a metaprogramming way, which may be key for performance.
Future work is also required to optimize the transformations and to speed-up the execution of metaprograms.

\balance
\bibliographystyle{abbrv}
\bibliography{references}

\end{document}